\documentclass[letterpaper,twocolumn,10pt,fullpage]{article}

\usepackage{times}
\usepackage{cite}	% sort citation numbers
\usepackage[hyphens]{url}
\usepackage[colorlinks,linkcolor=blue,citecolor=blue,urlcolor=black,breaklinks]{hyperref}
\usepackage{xspace}
\usepackage{graphicx}
\usepackage{paralist}
\usepackage[capitalize]{cleveref}
\usepackage{breakurl}

\long\def\com#1{}
\long\def\xxx#1{{\color{red} {\bf XXX }{\small [#1]}}}
\long\def\xxx#1{}
\newcommand{\ie}{{\em i.e.}\xspace}
\newcommand{\eg}{{\em e.g.}\xspace}

\newcommand{\trot}[2]{\rlap{\rotatebox{#1}{#2}~}}

\newcommand{\ty}{{\color{green}{$\surd$}}}
\newcommand{\tn}{{\color{red}{-}}}
\newcommand{\tq}{{\color{yellow}{?}}}

\begin{document}

%\title{Digital Personhood Online: Privacy, Transparency, and Inclusion \\
%	for Digital Democracy}
\title{Identity and Personhood in Digital Democracy: \\
	Evaluating Inclusion, Equality, Security, and Privacy in \\
	Pseudonym Parties and Other Proofs of Personhood}

\author{Bryan Ford \\
	Swiss Federal Institute of Technology in Lausanne (EPFL)}

\maketitle

\begin{abstract}
Digital identity seems at first like a prerequisite for digital democracy:
how can we ensure ``one person, one vote'' online without identifying voters?
But the full gamut of digital identity solutions --
\eg, online ID checking,
biometrics,
self-sovereign identity,
and social/trust networks --
all present severe flaws in security, privacy, and transparency,
leaving users vulnerable to exclusion, identity loss or theft,
and coercion.
These flaws may be insurmountable because
digital identity is a cart pulling the horse.
We cannot achieve digital identity
secure enough to support the weight of digital democracy,
until we can build it on a solid foundation of \emph{digital personhood}
meeting key requirements.
While identity is
about distinguishing one person from another through attributes or affiliations,
personhood is about giving \emph{all} real people inalienable digital
participation rights independent of identity,
including protection against erosion of their democratic rights
through identity loss, theft, coercion, or fakery.

We explore and analyze
alternative approaches to \emph{proof of personhood}
that might provide this missing foundation.
\emph{Pseudonym parties}
marry the transparency
of periodic physical-world roll-call events
with the convenience of digital tokens between events.
These tokens represent limited-term but renewable digital personhood claims,
usable for purposes such as
online voting or liquid democracy,
sampled juries or deliberative polls,
abuse-resistant social communication,
or minting universal basic income in a permissionless cryptocurrency.
Enhancing pseudonym parties to provide participants
a moment of enforced physical security and privacy
can address the coercion and vote-buying risks
that plague today's E-voting and postal voting systems alike.
We also examine other recently-proposed approaches to proof of personhood,
some of which offer conveniences such as all-online participation.
These alternatives currently fall short of
satisfying all the key digital personhood goals, unfortunately,
but offer valuable insights into the challenges we face.

\com{
Participants attend periodic real-world meetups
and obtain digital roll-call tokens,
which attest their unique personhood as demonstrated by their physical presence,
but need not reveal or prove anything about their identity.
These tokens represent limited-term but renewable rights
to digital ``citizenship'' in a democratic community.
We expect tokens to be usable for purposes such as public or private
digital communication, online voting or liquid democracy, deliberative polling,
information filtering, or minting community cryptocurrencies.
Federated pseudonym parties can scale securely and transparently
via simultaneous roll-calls at multiple -- perhaps many -- locations.
By offering participants a moment of
physical security and privacy, the roll-call process can address the coercion
and vote-buying risks that plague current E-voting and postal voting systems
alike.
Because participants need not show identity or prove anything to
participate, only loss of physical freedom or life itself can lead to
exclusion.  We expect to address physical mobility limitations and other
challenges in ongoing work.
}

\end{abstract}

\tableofcontents

\section{Introduction}
\label{sec:intro}

Who governs our digital world, and on what foundations?
Who decides what is allowed speech in online forums,
what is real or fake news,
who is a legitimate expert on a topic and who is a charlatan,
and ultimately how our online world will evolve?
We are presently faced with only fundamentally flawed, undemocratic answers.
Most governments do not wish to take on governance of the digital ecosystem,
rightly perceiving this to be outside their expertise
and at high risk of stifling innovation if they tried.
Even democratic governments in any case
represent the wrong constituency:
a government's jurisdiction is defined by a geographic border,
while online communities are geographically borderless.
But the alternative answers are just as bad.
Governments and public demands alike are rapidly delegating most
online governance power to unelected and unaccountable tech companies,
forcing on them a vague mandate to ``deal with''
hate speech and fake news and so on,
without providing even a hint of a plan for
how this opaque governance by algorithmic and employee decisions
might be made accountable to, let alone transparent to,
the online communities being governed.
Finally, the largely self-selected volunteers moderating and governing
popular sites like wikipedia and reddit may be largely well-intentioned
and competent within their interest areas,
but remain unrepresentative of and unaccountable
to the larger online public they govern,
often dominated by first-comers and loudmouths,
and prone to splinter into polarized in-groups and factions,
many of these contributing to the ongoing rise of violent extremism worldwide.
In short, what is the foundation we are clearly missing
for digital democracy and accountable, transparent governance
in the online world?

The most basic element of this conundrum is the question:
\emph{who} even is a legitimate member of an online democratic constituency,
who should wield a vote in governing an online community --
and how do we ensure that each such constituent wields only \emph{one} vote,
given how easy it is to create many fake account
with stolen or algorithmically synthesized online identities?
It is widely presumed that digital identity is a --
perhaps ``the'' --
key to placing digital democracy on a secure footing.
This paper begs to differ,
proposing instead that digital identity
is neither necessary nor sufficient for digital democracy,
but is rather a corrosive distraction.
Digital identity focuses on digitizing and verifying \emph{attributes}
that distinguish between -- and effectively divide -- people:
name, gender, origin, nationality, race,
education, certificates, wealth, connections, achievements.
The basic principle of equality implies that such attributes
should be irrelevant to participation in democratic processes.
Using attributes to \emph{identify} people digitally
only compromises the inclusion, equality, security, and privacy goals
of our presently-faltering attempts at founding any real digital democracy.

Instead, the central missing foundation that digital democracy needs is 
\emph{digital personhood}:
an enforceable assurance that
every real, natural human person may participate freely in digital democracy,
expressing their true and uncoerced preferences
in online governance,
while exercising one and only one vote
in online agenda-setting, deliberation, and decision-making.
Any attribute-focused identity -- including any digital identity --
may be lost, stolen, purchased, or misused coercively.
Digital personhood, in contrast,
is inalienable in the way our bodies are,
and in the way we take fundamental human rights to be.
The only way digital personhood can be lost
is by death or permanent incapacitation,
no matter what a person's identity attributes might be
or how their ability to prove them might change.
Recognizing that digital personhood rather than digital identity
is the most essential missing foundation for digital democracy, then,
we need some mechanism or process --
a \emph{proof-of-personhood}~\cite{borge17PoP} --
to validate and protect the digital personhood
of every real, human participant in an online community.

What fundamental characteristics of digital personhood
must a proof-of-personhood mechanism take into consideration,
in order to satisfy the needs of digital democracy?
We explore four:
proof-of-personhood must be
inclusive, equal, secure, and private.
Proof-of-personhood must be inclusive of nearly every
living, able-bodied person wishing to participate,
independent of factors such as
nationality, wealth, race, gender, connections, education, or expertise.
Proof-of-personhood must further ensure that each participant
obtains an \emph{equal} fundamental basis
for participation in digital democracy:
``one person, one vote.''
Proof-of-personhood must protect individuals from misuse
of their digital devices and credentials,
and must protect the democratic collective against subversion
through digital identity forgery, astroturfing, social bots~\cite{shao18spread},
and other Sybil attacks~\cite{douceur02sybil}.
Finally,
proof-of-personhood must protect individuals' privacy
to ensure that they can freely exchange information
and express their true preferences in digital deliberation and voting,
free from corrupt and undemocratic influence
through surveillance, coercion, or bribery.
\com{
Finally, digital personhood must scale to include
all natural persons wishing to (and sufficiently free to)
participate worldwide,
without needing support or permission from existing government jurisdictions --
though welcoming and leveraging government support where available.
}

This paper explores potential proof-of-personhood mechanisms,
starting by expanding on the original proposal of
\emph{pseudonym parties}~\cite{ford08nyms}.
Pseudonym parties are periodic real-world events
where people wishing to wield a vote online
gather to demonstrate their genuine personhood publicly,
each obtaining one-per-person digital tokens usable
for voting and other purposes during the next time period.
Pseudonym parties rely for security on the fact that
real people still (at present) have only one physical body each
that can be in only one place at a time.
As such, pseudonym parties embody no requirement
that people be \emph{identified} in any way,
or have any wealth, status, or connections, in order to participate.
This paper builds on prior explorations of pseudonym parties
as a proof-of-personhood foundation
by addressing further challenges such as
securely scaling to large (even global) federations of pseudonym parties
without falling prey to digital fakery attacks by malicious organizers,
inclusion of people who wish to participate but cannot due to timing,
and the challenge of minting coercion-resistant voting tokens
in pseudonym parties.

We then turn to other alternative approaches to proof-of-personhood
that have been proposed more recently~\cite{siddarth20who},
analyzing some of their strengths and weaknesses briefly and informally,
while deferring a detailed and rigorous analysis to future work.
Biometrics, for example,
represents an approach whose usability and scalability has been amply proven
by its deployment to over a billion users in India~\cite{abraham18state}.
Closer inspection and practical experience, however,
reveals tremendous security and privacy risks as well as inclusion failures.
Government-based identity and self-sovereign identity~\cite{allen16path}
similarly focus mistakenly on identity rather than personhood,
yielding privacy-invasive mechanisms that still
cannot adequately protect online democracy against large-scale digital fakery.
Investment-based foundations for decentralized permissionless cryptocurrencies,
such as Proof-of-Work~\cite{dwork92pricing,jakobsson99proofs,nakamoto08bitcoin}
and Proof-of-Stake~\cite{kiayias16ouroboros,gilad17algorand},
are prone to ``rich get richer'' effects progressively concentrating power,
and in any case fail to satisfy the equality principle of democracy.
Proof-of-personhood based on social trust networks~\cite{shahaf20genuine}
presume that people ``know'' and ``vouch for'' each other,
and in some cases even verify each other's humanness online or in-person.
But such verification mechanisms are necessarily
privacy-invasive and exclusionary if they are strong enough to work at all.
Even when working, they fail to protect against gradual accumulations
of false identities through interactions with disjoint subsets of real people.

In conclusion we find that, at present anyway,
federated pseudonym parties appear to be the only plausible means
to satisfy and strongly protect
all four critical properties of digital personhood --
inclusion, equality, security, and privacy --
and to lay a borderless, permissionless foundation
for genuine, truly representative digital democracy.
Nevertheless, there are remain many challenges and open questions
about how to make proof of personhood mechanisms both secure and usable.

\section{Goals for Digital Personhood}
\label{sec:goals}

If digital personhood represents the foundation
of a comprehensive architecture
for digital democracy~\cite{ford20technologizing},
what requirements would this base layer need to fulfill?
\com{
We first explore what properties a digital personhood mechanism
would need to have in order to provide
a truly solid foundation for participation in digital democracy.
}
In brief,
digital personhood should be:
\begin{itemize}
\item	{\bf Inclusive:}
	Any real human person should be able to participate,
	regardless of nationality, wealth, race, gender, connections,
	education, or expertise.

\item	{\bf Equal:}
	All participants must be treated equally
	for democratic deliberation and decision-making purposes:
	\ie, ``one person, one vote.''

\item	{\bf Secure:}
	Digital personhood must protect both individuals
	and the democratic collective from compromise
	in the digital and physical domains.

\item	{\bf Private:}
	Digital personhood must guarantee each participant's
	freedom to communicate, associate,
	and express their true intent in democratic processes.

\com{
\item	{\bf Scalable:}
	Digital personhood must scale to support
	everyone willing and able to participate,
	up to the world's current and future population.
}
\end{itemize}

We next develop and unpack these goals below.

\paragraph{Inclusive:}
Participation must not depend on race, gender,
or other personal attributes.
Participation must not depend on nationality or citizenship,
which excludes the stateless and many refugees and others
who find themselves unable to prove their citizenship.
Participation must not depend on wealth or related privileges
such as social connections or education,
or even on having one's own digital device.
Participants must not have to be technology-savvy,
to understand and follow complex rituals,
or to solve puzzles like CAPTCHAs or online Turning tests.
Most importantly,
digital personhood must be truly \emph{inalienable},
in all the ways that (digital) identity is not,
in that digital tokens and devices may be lost, stolen, sold, or forged
in the same ways that paper identity documents are.
Only death or permanent incapacitation
should be able to deprive a person of digital personhood
or participation in digital democracy.
A person who is \emph{not} permanently incapacitated
(and hence unable to contribute to society)
should always have a straightforward and accessible way
to recover and rebuild their digital personhood ``from scratch''
after any mishap in the real or digital world,
including complete a complete loss of documented identity, assets,
and even memory (\eg, amnesia).

\com{
\paragraph{Decentralized:}
No reliance on any central authority.
XXX part of inclusive?

\paragraph{Borderless:}
First-class participation not confined
to particular geographic jurisdictions.
Not dependent on the permission or support of traditional governments,
while welcoming their participation and support whenever available.
XXX part of inclusive?
}

\paragraph{Equal:}
All participants should have equal foundational power and influence
in democratic decision-making.
It must not be realistically feasible for any individual or organization
to buy more (effective) voting power
with more wealth, connections, or other resources.
This requirement excludes many common digital identity proxies
that are convenient and commonly-used but readily purchasable,
such as phone numbers, IP addresses, or credit card numbers.
It also excludes investment-based foundations for decentralized systems
such as proof of work or proof of stake.

\paragraph{Secure:}
Digital personhood must protect individuals and the democratic collective alike
from abuse and subversion in the digital and physical domains.
Individuals must have strong protections against their digital devices
and credentials being misused or misappropriated by others.
The approach must securely ensure that only real, natural persons participate,
each wielding only one vote,
and must thereby securely exclude non-human digital entities
such as fake digital identities,
corporate astroturfing,
social bots,
and other forms of Sybil attacks~\cite{douceur02sybil}.
The security of digital personhood must be \emph{resilient},
surviving any single or threshold number of failures or compromises
in all security-critical architecture roles, human and digital alike.
Individuals must have inclusive paths to recover or rebuild
their digital lives even after the most extreme
physical or digital compromises.

\paragraph{Private:}
Digital personhood must protect each individual's privacy,
including each person's ability to communicate and associate freely
and express their true intent in democratic processes.
This includes protection from the use of digital devices or tokens
under duress, coercion, or bribery of any kind.
Almost all current approaches to E-voting
fail to satisfy this {\em coercion-resistance} requirement,
which is one key reason most voting security/privacy experts still
recommend in-person paper-based voting
and against any form of E-voting.
But it is a challenge we must confront and solve,
and not just give up on as ``too hard.''

\com{
\paragraph{Scalable:}
Digital personhood should be available to 
Should scale in principle to support the entire world's population.

XXX part of inclusive?
}

\xxx{ digital personhood, age, and consent: what about kids? }

We next explore proposed ways to implement digital personhood
through a \emph{proof-of-personhood} mechanism in some form,
starting with pseudonym parties.

\section{Pseudonym Parties}
\label{sec:pop}

This section first outlines pseudonym parties
as originally proposed~\cite{ford08nyms},
then informally analyzes this approach
against the goals outlined above.

\subsection{The basic idea}

In brief,
a pseudonym party gives each attendee at an in-person event
exactly one anonymous digital \emph{proof-of-personhood token}
or \emph{PoP token}.
The process is organized so as to leverage physical security,
and the fact that real people have only one body each,
to ensure that each person gets only one token.
Pseudonym parties are intended to be held periodically,
such as once a week, month, quarter, or year,
with the tokens minted at each party having a limited valid lifetime
only until the next periodic event.

After each event,
the organizers publish a list of the anonymous tokens they handed out.
Anyone can subsequently verify that the length of the published list
matches the number of actual attendees --
according to the direct observations of attendees themselves,
indirect reports from eyewitness observers,
and/or other evidence such as photos and videos
taken at the event and published.

A pseudonym party thus allows attendees to \emph{prove their personhood}
transparently in a public ceremony,
demonstrating their existence as a human
and obtaining a limited-term digital proof of their unique personhood,
without having to be \emph{identified} at all, by anyone.

\subsection{Protecting security and equality}

The main operational security objective in a pseudonym party
is ensuring:
(a) that each attendee obtains one digital token that only they control,
such as by holding a cryptographic private key for it;
(b) that each attendee obtains \emph{only one} such token,
to guard equality; and
(c) that the total number of people who were verifiably in attendance
closely if not exactly matches the length
of the list published after the event,
to ensure that corrupt organizers cannot manufacture fake virtual attendees
and reap the corresponding benefits for themselves.

Objectives (b) and (c) are most critical to collective security:
an attendee or organizer who can improperly obtain many tokens
may gain the power of all manner of Sybil attacks against the community,
including ballot stuffing,
sockpuppetry and astroturfing via many false identities,
obtaining many shares of community benefits such as basic income, etc.
To achieve these collective security goals, therefore,
a pseudonym party needs to be publicly transparent enough,
and documented through enough independent sources of evidence
both human and digital,
that this body of evidence and testimony can leave no reasonable doubt
in either the fact that the event occurred
or the number of people who attended.
Adequate measures to ensure such security and transparency
will naturally vary depend on conditions and an event's size.

\subsubsection{Small events}
At small events involving at most a few tens of attendees,
a simple and informal process may suffice for security,
because all attendees can simply watch and verify the process for themselves.
For example,
each attendee might use a PoP wallet app
to create a single-use token and display it as a QR code.
One designated organizer simply scans all of these QR codes
and broadcasts a list of them locally to all attendees.
The attendees check that the length of the broadcast list
matches their direct observation of the number of attendees,
and if so,
\emph{witness cosigns} the list as an eyewitness
attesting that fact~\cite{syta16keeping}.
Any attendee who sees that the organizer's list is longer
than the number of people physically present,
or sees the organizer scanning some attendee more than once,
or finds her own token missing from the final list,
can publicly complain, refuse to cosign,
and ultimately attend a different pseudonym party next time
if the problem is not rectified.

\subsubsection{Medium-size events}
If a pseudonym party graduates from a few tens to hundreds of attendees,
it becomes difficult for all attendees watching an organizer
either to count reliably,
or to remember which other attendees
the organizer has and has not yet counted.
This scale thus calls for a more structured process.
All attendees are asked to gather in an enclosed or demarked space,
such as a designated room or a cordoned-off outdoor area,
before a designated deadline.
This space might serve as a lobby in which to gather and socialize
before a ``main event'' commences,
such as a keynote speech or concert for example.
At the critical deadline -- when the main event is about to start --
all entrances to the lobby are closed so no one else may enter.
Those already in the lobby then move one at a time from the lobby area
to the main event area,
each attendee presenting a token QR code for scanning as they exit the lobby.
The organizer scans only one QR code per person,
publicly displaying the list and running count throughout the process.
Anyone present can watch and film both the scanning process
and the lobby as a whole,
both to convince themselves and provide eyewitness testimony if needed
that each attendee was scanned only once,
that no one was allowed to enter or re-enter the lobby after the deadline,
and that the number of attendees scanned matches
the length of the list subsequently published.

\subsubsection{Large events}
If a pseudonym party reaches thousands or more attendees --
as might happen in an event doubling as a political rally or protest,
for example --
then the basis for security remains the same but just needs to scale.
Standard crowd control measures
of the kinds commonly used in theme parks or to manage large protests --
such as portable barriers and officials watching them --
may apply in this case.
Instead of one organizer,
several or many organizers might scan attendees leaving the enclosed area
via multiple lines in parallel,
in order to accommodate large numbers of people
without causing inordinate wait times.
Many witnesses, both officially-designated and unofficial volunteers,
might film and publish video documentation of the event from all perspectives,
both broad and focused especially on the token scanning lines
and ``do-not-cross'' boundaries.
A sufficient number of guards must monitor the boundaries and exits,
and must have authority to catch and eject anyone attempting to (re-)enter
the critical enclosed space after the deadline until the event has closed.

\subsubsection{Federated pseudonym parties}
\label{sec:pop:fed}

Scaling beyond one geographic location securely
requires multiple groups to federate
to organize simultaneous events at multiple different locations.
The most basic security requirement in this case
is that all such federated events have synchronized entry deadlines.
That is, all events must close the entrances to their lobby areas
at the same time,
before starting to scan tokens,
so that it is impossible for any single person (with only one body)
to be present at and get tokens scanned at more than one such federated event,
even if they had instantaneous travel.

\paragraph{The timezone challenge}

There is in principle no barrier against such a federation
scaling to support regular simultaneous events
at every city, town, and village in the world.
The synchronization requirement does present a convenience problem
due to timezones, however:
a pleasant high-noon entry deadline in one place
is inevitably a 3AM deadline somewhere else on the globe.

If the globally-synchronized deadline varies from one event to the next,
however,
we can ensure that everyone will have the opportunity to attend
some events at a time convenient to them,
even if we expect most people to attend only a subset of events.
An alternative approach would be to divide the globe into, say,
three large federations encompassing about eight timezones each,
and accepting that a few determined people close to federation borders
will readily be able to obtain tokens (and corresponding benefits)
from the two adjacent federations.

\paragraph{The deep fake challenge}

The main remaining security challenge is keeping the group of organizers
at each location accountable and transparent to all the organizing groups
at other locations.
A key threat is that a corrupt group of organizers
might attempt to fabricate an event that did not occur at all,
or inflate their event's attendance,
using today's advanced graphics and in particular deep fake~\cite{chesney18deep}
technologies to forge a body of digital audio/video ``evidence'' convincingly.

Probably the strongest measure to mitigate such threats
is to ensure constant interaction and \emph{cross-witnessing}
between locations.
The organizers at any location should, in effect, be dead-certain that
their event is being observed, recorded, and publicly reported on
by multiple official and unofficial (volunteer) witnesses
who normally attend events at other locations,
and that any discrepancy between their claims and those witnesses'
and testimony will quickly be noticed and investigated.

Some such cross-witnessing can be expected to happen opportunistically
as a result of normal travel:
\eg, a person looking up and dropping in on a local event
during a business trip or vacation.
To ensure proactively that \emph{all} locations can anticipate 
some cross-witnessing at each event,
however,
and hence a high assurance of fabrication by any group being caught,
the federation might run a \emph{secret cross-witness} travel lottery.
Anyone normally attending some location can sign up,
and a subset of such volunteers are randomly selected
and secretly asked to travel to and serve to cross-witness
another randomly-assigned pseudonym party location in the near future.
Volunteers are offered modest compensation
for accepting their random cross-witnessing assignments,
depending on required travel distance.
Assigned cross-witnesses
are required to keep their status secret until after the assigned event.
At this point they can reveal (and prove)
their official cross-witness status after-the-fact,
together with the body of video and other evidence they recorded,
and their personal testimony on whether they thought the event was run properly
and any irregularities they might have observed.
A corrupt group of organizers would thus not only have to produce
a considerable body of convincingly-fake evidence
from many (fake) perspectives,
but also successfully bribe or coerce nearly every attendee
they don't recognize who shows up and could be a secret cross-witness.

\subsection{Privacy in pseudonym parties}

Pseudonym parties guard attendees' privacy
by not requiring them to show any form of ID
or submit to biometric tests.
The digital tokens scanned and published at each event
are merely cryptographic random numbers
that contain no personal information or traceable link to their owners.
Attendees might even wear masks and costumes at the pseudonym party,
as in a Venetian carnival,
to conceal even the fact of their attendance
from anyone who might recognize them.

\subsubsection{Privacy in the use of PoP tokens}

With appropriate design,
each attendee's subsequent \emph{uses} of their tokens
can also be cryptographically unlinkable from each other
and from the attendee's position on the published list,
ensuring strong privacy even for attendees
who did not wear a mask and might be known to have
received a particular token on the published list.

Pseudonym parties thus satisfy our main privacy goal for proof of personhood
by not collecting personally identifiable information (PII) in the first place,
and by cryptographically de-linking all subsequent uses
of the tokens from the tokens themselves.

\subsubsection{Coercion resistance}
\label{sec:pop:coercion}

A far more technically difficult privacy challenge
that is crucial to digital democracy
is \emph{coercion resistance}:
ensuring that a PoP token is used only by the intended person
under their free will and genuine consent.
One important use-case for PoP tokens is for online voting and deliberation,
and hence the well-known and extremely difficult
coercion-resistance challenges that E-voting systems face
translate into corresponding challenges for pseudonym parties as well.
For example, how can we prevent a person or organization with means
from secretly hiring or bribing many real people to attend a pseudonym party,
obtain one (legitimate) PoP token each,
and then use their respective tokens to vote in the interests of the coercer?

In today's digital ecosystem it is difficult to imagine a way
to detect or prevent such transactions from occurring
without precisely the kind constant privacy-invasive surveillance
we seek to avoid.
It is doubtful even that we can plausibly detect, track down,
and halt such attempts at coercion or vote-buying,
especially given that they can potentially be launched
from anywhere in the world,
such as in a country from which the perpetrator is unlikely to be extradited
even if caught.
The perpetrator might even launch the attack anonymously
via smart contract mechanisms such as dark DAOs~\cite{daian18on-chain},
leaving effectively no trace or link back to the perpetrator
once funded anonymously with cryptocurrency and launched.

Fortunately there is an alternative to the unrealistic prospect
of tracking down and deterring attempts at coercion.
Instead, we can protect the free will of pseudonym party attendees
by ensuring that even if they are bribed or otherwise coerced,
they need not ``stay bought.''
In particular, we can adapt to pseudonym parties
an approach to coercion resistance developed for E-voting,
in which each voter can obtain both \emph{real}
and \emph{fake} tokens~\cite{juels10coercion}.
The voter uses their real tokens secretly to vote
according to their own genuine interests,
while giving the fake tokens to anyone offering to buy their vote.
Only votes cast using real tokens actually count,
and only the voter who received the tokens knows which is which.

Coercion resistance generally requires voters to have
at least a moment of (genuine) privacy
despite being at a highly public ceremony --
hence the curtained privacy booths that are standard for in-person voting.
To achieve coercion resistance for pseudonym parties,
attendees similarly need a moment of privacy at a public event,
outside the control and surveillance of a potential coercer or vote-buyer.
In this case, the moment of privacy is not for the act of voting itself,
but instead to allow each attendee to obtain real and fake PoP tokens,
and to learn which is which during that moment of privacy.
Attendees must then \emph{know} but be \emph{unable to prove} this fact
after the moment of privacy ends,
once the attendee may again be subject to coercion.
The coercer thus has no way to verify
whether the attendee complied and hence ``stayed bought.''

One way we might implement coercion resistance
at pseudonym parties is as follows.
As each attendee exits the lobby,
instead of getting a token scanned immediately,
the attendee instead receives a single-use ticket from the organizer
managing that exit line.
The attendee then deposits
any recording-capable electronic devices temporarily
at a check-in desk,
then enters one of several curtained privacy booths.\footnote{
	The requirement to check electronic devices is to ensure that
	attendees cannot be successfully bribed or otherwise coerced
	to compromise their own privacy
	by recording or live-streaming their activities in the privacy booth,
	for example.
	The requirement to check electronic devices might be enforced
	by a metal detector if the threat is sufficiently severe.}
The attendee inserts the ticket into
a kiosk in the privacy both,
which prints one real token and several fake tokens on paper.
The attendee knows that the first token printed is the real token,
and may mark the printed tokens as an aid to remembering which is which.
Upon leaving the privacy booth, however,
\emph{only} the attendee knows which is the real token,
and cannot subsequently prove which is which to anyone else.

In the example of a coercer who hires people
to attend a pseudonym party and vote in the coercer's interest,
the attendees can safely give the coercer their fake tokens --
claiming unfalsifiably that they are real --
while in fact double-crossing the coercer
by using their real tokens to vote in their own interests
and not the coercer's.
The design of this coercion-resistant token-printing process and kiosk
presents technical and security/privacy challenges, of course,
which is currently a work-in-progress.
The bottom line, however, is that these challenges appear solvable
without attendees having to trust the kiosk
for anything \emph{other than} coercion resistance.
Even a fully-compromised kiosk cannot undetectably forge Sybil tokens
or steal the tokens of uncompromised, uncoerced attendees.

In extreme cases such as domestic coercion,
an abusive partner or relative
might lurk nearby at the pseudonym party itself,
monitoring the victim at every moment
in line and as they enter the privacy booth,
then again after they emerge from it.
In this case the victim may have no safe means
of leaving with their real token and using it elsewhere secretly.
One way of addressing this extreme coercion case
is to enable the attendee to use their real token
\emph{in the privacy booth}
to delegate their subsequent normally-online votes to a party of their choice,
as is already common in party-list proportional-representation elections,
or even to delegate to an arbitrary friend they trust to represent them
as in liquid-democracy systems~\cite{blum16liquid,ford20liquid}.
The coercee then discards the real token in the booth
and leaves holding only a fake token,
which they can then present to their coercer
or use under the coercer's surveillance.
While this approach unfortunately eliminates the victim's opportunity
to participate online in more fine-grained deliberation and voting
between pseudonym party cycles,
it at least preserves their ability to express their free will
in relative safety.

With appropriate design, therefore,
pseudonym parties can potentially not only ensure
a secure ``one person, one vote'' distribution of tokens to real people,
but can also ensure that the people receiving those tokens
have the opportunity to use them under their own genuine free will,
even in the presence of resourceful coercers, either nearby or remote,
who may be unlikely to be caught or deterred.

\subsection{Inclusion}
\label{sec:pop:inclusion}

The fact that pseudonym parties need not collect or verify
any identity or biometric information
also addresses many inclusion challenges, though not all.
The proof-of-personhood (PoP) tokens handed out at a ceremony
are anonymous random numbers
that clearly encode no information about
an attendee's gender, race, wealth, nationality, or other characteristics.
If many attendees wear masked costumes and some regularly cross-dress,
these practices can head off risks of discrimination or exclusion at the event
on the basis physical characteristics such as race, age, or gender.

The main exclusion risk that pseudonym parties potentially have trouble with
concerns people who wish to attend but cannot due to
lack of mobility or freedom at the event's designated time.
Prisoners and residents of authoritarian surveillance states, for example,
may clearly be prevented from organizing or attending a pseudonym party.
It is hard to envision \emph{any} approach
successfully guaranteeing that a person can obtain and use
a proof-of-personhood token freely and privately without duress or coercion,
if that person is under constant surveillance
and hence by definition has no effective privacy or freedom.

Less-extreme scenarios are thus actually more worrisome in practice,
such as exclusion of those whose jobs require them to be on-duty elsewhere
at a pseudonym party's designated time and place.
Holding successive pseudonym parties at varying times and dates
may help ensure that even those with restrictive schedules
can participate in some events, if only a subset.
If pseodonym party attendance translate into
economic benefits such as a \emph{crypto-UBI}
or universal basic income
in cryptocurrency form~\cite{ford19money,zhang20popcoin},
then employees might argue for employers to reimburse them
for attendance benefits missed due to their work schedulers.

Another possibility may to allow some attendees in exceptional cases
to register before an event for ``absentee participation,''
and consent to verifiable location tracking \emph{during} the event,
publicly proving via multiple independently-verifiable forms of evidence --
such as via location tracking devices together with eyewitness attestations --
showing that they were at work and not attending any pseudonym party.
This approach could thus allow people otherwise excluded by responsibilities
to participate ``remotely'' at the cost of
a brief, opt-in compromise of their location privacy at the critical time.

\com{
mobility-reduced: wheelchairs OK,
but what about bedridden in hospitals?
}

\subsection{Pseudonym parties in pandemic times}

It is ironic to be writing a proposal for large in-person gatherings
during a global pandemic,
in which most large in-person gatherings
are forbidden for public health reasons across much of the globe.
We hope that the current situation is not permanent, of course.
But what if it is, and a ``new normal'' persists indefinitely
in which people must avoid dense gatherings especially indoors,
remain widely distanced even when outdoors, and so on?

There is nothing preventing us from organizing pseudonym parties
primarily outside,
to ensure ventilation and adequate distancing between attendees throughout.
The main challenge is reserving and cordoning off enough space
for the number of people expected to gather
in the ``lobby'' area before the deadline.
Large public parks might be used and painted with distancing circles
for attendees relaxing or socializing in the enclosed area,
as has already been done at parks in New York~\cite{harrouk20domino},
San Francisco~\cite{tyska20photos},
and other cities.
Clearly-marked areas might be reserved for those waiting in line
to obtain a token and leave the lobby area,
with the distance markers that have become standard for such lines.

One challenge is urban or suburban neighborhoods
without sufficiently-large parks nearby
to accommodate safely the number of people wishing to attend.
Those with cars might travel to more distant, larger spaces,
but expecting everyone in a dense area to do so
would be either unsafe or exclusionary to those
who would have to use shared public transportation for that travel.
Temporarily closing local neighborhood streets to vehicle traffic,
and using those in addition to or instead of park areas,
may be a less-comfortable but workable solution.

Weather is another important consideration, of course,
which will certainy affect peoples' willingness to participate
in pseudonym parties,
just as it already can affect
voter turnout in traditional elections~\cite{gomez07republicans}.
Scheduling yearly events in summertime may help --
but if such an event is synchronized globally
as discussed above in \cref{sec:pop:fed},
we face the problem that one hemisphere's summer
is the other hemisphere's winter.
Scheduling yearly or semiannual events in the spring or fall
may be a better compromise in this regard,
at least avoiding the coldest periods anywhere.

But we might desire
more frequent quarterly, monthly, or even weekly events,
to reduce the time newcomers must wait to obtain their first PoP token,
and to reduce the impact to regular attendees having to miss one cycle.
There will then be no escape from the chance of bad weather
in most parts of the world.
Appropriate architectural measures,
such as large temporary or permanent shelter structures
with high ceilings but open walls,
may help attendees maintain reasonable comfort
while gathering in a space with safe distancing and ventilation.

\subsection{Use cases for PoP tokens}
\label{sec:pop:uses}

Although this paper's focus is primarily
on ways to \emph{create} proofs of personhood securely
rather than on applications for them,
we briefly summarize a few promising use cases and how they might function.

\subsubsection{An alternative to CAPTCHAs}

Web sites often use automated Turing tests or CAPTCHAs~\cite{ahn03captcha}
to rate-limit automated abuse attempts,
such as miscreants attempting to create many fake accounts.
As machine-learning techniques have improved, however,
web sites have had to increase the difficulty of these CAPTCHAs progressively,
until real humans often have as much difficulty solving them
as machines do~\cite{dzieza19why}.
CAPTCHAs are often exclusionary to those with disabilities or language barriers,
and are annoying and time-consuming even to those who can usually solve them.

Web sites and online services of all kinds
could allow users to bypass CAPTCHAs automatically using a PoP token.
The online service can tell whether or not a particular PoP token
has already been used for a particular operation on that service,
such as signing up for a new account,
although this use of the PoP token reveals nothing else about its holder.
Because one real human user receives only one PoP token
each time he attends a pseudonym party,
PoP tokens offer online services much stronger rate-limiting protection
against automated abuse of their services than CAPTCHAs do.
An abuser can expect to get a new PoP token only once
a week, month, quarter, or year,
whereas either a human abuser or CAPTCHA-solving bot
can successfully solve a CAPTCHA in a matter of seconds.
And PoP tokens offer the user requesting the service the greater convenience
of immediate access without having to solve increasingly-difficult puzzles.

The PoP token itself need not be revealed to the online service
when used in lieu of CAPTCHA solving.
Instead, the service receives only a cryptographically-unlinkable \emph{tag}
that it knows has a 1-to-1 relationship to some valid PoP token,
even though no one but the token's owner knows which one.
One way to implement such a mechanism cryptographically
is using compact \emph{linkable ring signatures},
for example~\cite{au06short,tsang05short}.
When the user uses the same PoP token to access different online services,
those services obtain distinct and cryptographically-unlinkable tags,
which are therefore unusable to track users across different services.

\subsubsection{Verified likes and follower counts}

As soon as social media became a primary communication channel
and competitive field for information sharing and advertising,
unscrupulous fraudsters soon started synthesizing fake identities
or \emph{social bots} to promote particular viewpoints or content,
or to increase the apparent reputation and influence of a real account
by inflating its follower count~\cite{ferrara16rise,bessi16social}.
One way social media platforms could use PoP tokens from pseudonym parties
is to neutralize Sybil attacks from social bots,
by displaying and using only counts of \emph{unique real people}
when computing and display ``follower'' or ``like'' counts
or selecting items for a user's feed.

With such a measure properly implemented,
social media platforms need not forbid users
from creating multiple accounts for different purposes
or representing different sides to their personality --
\eg, a professional feed, a personal feed,
one for a favorite hobby, etc.
The policy Twitter exemplifies of even allowing
well-behaved bot accounts on the platform may be embraced,
since interacting with bots can sometimes be useful or entertaining.

But whenever any of a user's online accounts ``likes'' or upvotes a post,
that upvote gets counted in displayed statistics,
newsfeed selection, and other algorithms
\emph{only} if the account has a PoP token
valid at the time of the upvoted post.
Accounts whose users currently have no valid PoP token at that time --
\eg, because the owner missed the last pseudonym party cycle, or is a bot --
can still upvote items, but these upvotes have no impact
on aggregate statistics or content selection.
Similarly,
if a user upvotes the same post through multiple accounts he controls,
all of these upvotes count only once
because they are linked to the same PoP token.
Thus, each PoP token valid at the time a given post appears
serves as a single right for its holder's upvotes
to be counted once and only once.

Follower counts might similarly be computed in ``one-per-real-person'' fashion.
Since (real) accounts generally have extended lifetimes
rather than being of interest only at a particular moment in time,
platforms might simply use the currently-valid PoP tokens at any given time
in calculating follower counts to display on social media accounts.
Each follower of a given account is actually counted
only if that follower currently has a valid PoP token,
and is counted only once even if multiple follower accounts
use the same PoP token.
Thus, one's follower count might change not only as a result
of other accounts following or un-following them,
but also as a result of a follower obtaining and linking a PoP token,
or of a follower's PoP token expiring without being renewed.

%com

\subsubsection{Online voting and deliberation}

It almost goes without saying that PoP tokens
may be used in online voting and deliberative processes,
ensuring that each real person wields only one vote,
even if they have multiple online accounts representing different personas.
As soon as higher-stakes democratic processes move online,
ensuring coercion resistance becomes more critical,
as discussed earlier in \cref{sec:pop:coercion}.

One practical issue we may rightfully worry about 
is the effective disenfranchisement of anyone
who had to miss the last cycle of pseudonym parties,
at least until the next cycle that they manage to attend.
While this temporary loss of voting power is an important concern,
it is not fundamentally different or worse
than the effective disenfranchisement we have today
of voters who cannot readily make it
to a conventional election that normally requires in-person voting.
If pseudonym parties were to become a widely-used mechanism,
businesses and governments might hopefully establish policies
that help enable most people the freedom to be off-duty if desired
around the most important pseudonym party cycles.
For the cases in which this is impossible,
exception-case mechanisms of the kind discussed earlier
in \cref{sec:pop:inclusion} may apply.

Other innovations in digital democracy
might also ``soften the blow'' of temporary disenfranchisement
due to missing pseudonym party cycle.
In liquid democracy, for example,
eligible voters who do not wish to -- or have no time to --
follow all the details of an online discussion or deliberative process
can \emph{delegate} their vote temporarily
to a chosen representative~\cite{blum16liquid,ford20liquid}.
A person who \emph{does} have the time and interest 
in participating in the deliberation closely,
and acquires a reputation for being knowledgeable and trustworthy
on the topic of discussion,
may thus build up and wield a significant amount
of delegated proxy voting power.
If this respected authority must miss one pseudonym party cycle,
she temporarily loses only her single \emph{individual} vote,
and not the ability to wield the delegated voting power
of others who \emph{did} obtain a PoP token in the most recent cycle.
Thus, her voting weight merely drops by one vote until the next cycle,
rather than falling to zero.

\subsubsection{Sortition-based juries and deliberative polls}

Another important potential online governance structure
that PoP tokens could support is sortition-based selection
of juries, members of deliberative polls~\cite{fishkin05experimenting},
or other \emph{open democracy} processes
that might need to be diverse and representative
but manageable in size~\cite{landemore20open,landemore20reinventing}.

A government, organization,
or online association might use a recently-published list of PoP tokens
to ``call'' a randomly-selected sample of people
to participate in a deliberative poll or jury.
Even though these sampled PoP tokens are anonymous to everyone else,
their holders can tell which published PoP tokens are theirs.
The token holder's PoP wallet might notify the owner
if a call arrives for sortition-based participation
in a process of potential interest, for example.
The organizer can guarantee that the selection of called PoP tokens is fair
by relying on the output of a decentralized  random
beacon~\cite{syta17scalable}
such as {\tt drand}.\footnote{\url{https://drand.love}}
Because each PoP token represents exactly one human user
who attended some pseudonym party in the last cycle,
each real person gets an equal chance of selection
regardless of how many online accounts or identities they might have.

\com{
with or without coercion resistance...

liquid democracy

deliberative polling

\subsection{Accountable anonymity services}

Tor etc.
}

\subsection{Pseudonym parties wrap-up}

In conclusion,
while organizing and scaling pseudonym parties securely
presents numerous technical and logistical challenges,
this approach appears to present a clear path to achieving
the key goals of proof of personhood in a strong form:
inclusion, equality, security, and privacy.

\section{Alternative Approaches}
\label{sec:alts}

We now turn to examining and informally analyzing
a number of other approaches to proof of personhood
that have been proposed recently
and even prototyped~\cite{siddarth20who}.
A more detailed and rigorous analysis is left for future work. 
We commence by briefly classifying proof-of-personhood approaches
by key features,
then analyze these classes one-by-one,
examining unique features of individual approaches only as needed.

\subsection{Classifying alternative approaches}

\com{
\begin{table*}[t]
\begin{center}
\begin{tabular}{l|*7c}
Approach
	& \trot{60}{Inclusive}
		& \trot{60}{Equal}
			& \trot{60}{Usable}
				& \trot{60}{Secure}
					& \trot{60}{Private}
						& \trot{60}{Resilient}
							& \trot{60}{Scalable} \\
\hline
Government Identity
	& \tn	& \tq	& \tq	& \tq	& \tn	& \tn	& \ty	\\
Biometric Identity
	& \tq	& \ty	& \ty	& \tq	& \tn	& \tn	& \ty	\\
Self-Sovereign Identity
	& \ty	& \tq	& \ty	& \ty	& \tn	& \tn	& \ty	\\
Proof-of-Work
	& \tq	& \tn	& \ty	& \tq	& \ty	& \ty	& \ty	\\
Proof-of-Stake
	& \ty	& \tn	& \ty	& \ty	& \ty	& \ty	& \ty	\\
Social Trust Networks
	& \tn	& \tq	& \tn	& \tn	& \tn	& \tq	& \tq	\\
Online Verification
	& \tq	& \tq	& \tn	& \tq	& \tq	& \ty	& \ty	\\
Federated Pseudonym Parties
	& \ty	& \ty	& \ty	& \ty	& \ty	& \ty	& \ty	\\
\end{tabular}
\caption{Summary of some alternative approaches
	to digital identity and personhood}
\label{tab:alts}
\end{center}
\end{table*}
}%com

\begin{table}[t]
\begin{center}
\begin{tabular}{l|*7c}
Approach
	& \trot{60}{Inclusive}
		& \trot{60}{Equal}
			& \trot{60}{Secure}
				& \trot{60}{Private} \\
\hline
Government Identity
	& \tn	& \tq	& \tq	& \tn	\\
Biometric Identity
	& \tq	& \ty	& \tq	& \tn	\\
Self-Sovereign Identity
	& \tq	& \tq	& \ty	& \tn	\\
Proof of Investment
	& \ty	& \tn	& \ty	& \ty	\\
Social Trust Networks
	& \tn	& \tq	& \tn	& \tn	\\
Threshold Verification
	& \tq	& \tn	& \tq	& \tq	\\
Pseudonym Parties
	& \ty	& \ty	& \ty	& \ty	\\
\end{tabular}
\caption{Classification and analysis summary
	of several alternative approaches
	to digital identity and personhood.}
\label{tab:alts}
\end{center}
\end{table}

To provide a broad comparison of alternatives,
we examine not only approaches that explicitly set out
to solve the ``unique human'' or proof of personhood problem,
but also other Sybil-resistance mechanisms
that are widely-known and commonly-used to achieve overlapping
if not identical goals.
\Cref{tab:alts} concisely summarizes this classification.
For each broad approach
and each of the four key proof-of-personhood goals,
the table also summarizes whether our informal analysis finds
the goal to be satisfiable with strong confidence (\ty),
only questionably satisfiable (\tq),
or definitely unsatisfied (\tn),
for reasons elaborated further below.

Analyzing approaches to proof of personhood
and closely-related Sybil-resistance schemes this broadly
presents numerous challenges, of course.
Many proposed approaches have limited (and usually not peer reviewed)
documentation on how the approach actually works,
and often lack a well-defined threat model
or statement of goals and assumptions.
Our analysis will therefore be subjective in many ways,
and necessarily but admittedly unfair
in that we will analyze proposed approaches ultimately
against \emph{our} threat model and assumptions --
particularly the four proof of personhood goals
set out in \cref{sec:goals} --
and not the (usually-implicit) goals and assumptions
the authors may have intended.

In hopes of compensating at least partially for this unavoidable unfairness
in comparing proposed approaches to what might be viewed as our ideal standard,
we attempt to give each approach the benefit of the doubt
by evaluating the properties that each broad class of approaches
appears likely \emph{capable} of satisfying with appropriate design.
For this reason, some cells in \cref{tab:alts} are marked satisfiable (\ty)
even in cases where it is unclear and perhaps doubtful
that current specific approaches \emph{do} satisfy those properties,
when there appears to be a clear path toward filling those gaps.
\Cref{tab:alts} therefore marks a property unsatisfied (\tn)
only where there appears to be a fundamental reason
that class of approaches \emph{cannot} satisfy the given property,
without modifying basic premises
and hence becoming a different approach entirely.

Several of the approaches we explore below
make no pretense at achieving the goals of proof of personhood \emph{per se},
but we nevertheless examine them for broad comparison purposes.
Readers interested only in schemes that specifically attempt
to address the ``unique human'' problem central to proof of personhood
may wish to skip ahead to \cref{sec:alts:social}.

\subsection{Government-issued identity}
\label{sec:alts:govt-issued}

As a natural comparison baseline,
we first briefly consider the approach most governments today
use to verify an individual's personhood
and eligibility to receive the benefits of government services:
that is, identity documents, either paper-based or increasingly digital,
that attest to a person's origin (\eg, birthdate and birthplace),
status such as citizenship or residency,
and identifying personal characteristics such as photo or other biometrics.
Based on historical experience,
it seems questionable at best whether government-issued identity
can robustly satisfy any of the four goals of digital personhood.

When a person uses an identity document
(\eg, a drivers license or passport)
to prove their eligibility for some benefit
(\eg, entering a country or receiving unemployment or social security benefits),
verification is based entirely on matching
the person's physical characteristics and/or knowledge
against existing documents or databases.
For example, an ID checker -- either human or increasingly machine --
typically compares the person's face
and sometimes other biometrics like fingerprints
against those on a photo ID or in an electronic database,
and sometimes asks the person questions about their past
such as mother's maiden name, birthplace,
a recent bank transaction, favorite pet, etc.
Even when signing up for or renewing an identity document such as a passport,
the process generally relies for security on similar verification
of \emph{other} earlier identity documents the person already had:
\eg, a birth certificate, social security card,
earlier expired passport or other ID, etc.

All of these documents tend to be forgeable in practice at varying costs,
and the security of their issuance generally has numerous
single points of compromise.
For example, to start building a false identity around
a name from a gravestone or a stolen profile,
a determined identity fraudster often need only find
a single human office worker who can be confused through social engineering,
or successfully bribed or extorted,
to accept weak, synthesized or tampered-with evidence of the person's past
such as forged birth certificates and other papers.
The fact that identity documents may be and often are lost or stolen,
for many legitimate reasons,
forces governments to have processes essentially relying on a person's say-so
to re-established documented status,
and these processes similarly present opportunities for fraudsters to exploit.

Further, there is essentially nothing about one documented history of a person
guaranteeing that \emph{another} documented history
of the same physical person --
or perhaps many such histories --
cannot exist or is likely to be discovered if they do exist,
unless the multiple-identity fraudster simply makes an egregious mistake.
Criminal organizations and spy agencies alike rely on this fact routinely,
and when their members are caught using false identities,
it is often due in part to mistakes made
in using those false identities while keeping them separate.
\xxx{ citations }

\xxx{ digitalization: blockchain and AI }

Thus, it is questionable at best whether the government identity approach
can guarantee either security with no single point of failure,
or equality in ensuring that each person obtains only one identity,
without subjecting people to far-more-invasive scrutiny
and verification processes than liberal democracies tolerate at present.
For example, to ensure true threshold security
with no single points of compromise,
everyone applying for or renewing a drivers license or passport
might ultimately need not only to interact with one government office worker,
but rather to convince a multi-member ``identity inquisition committee''
that the applicant's existing documentation and claimed history is legitimate.

These issues ultimately highlight the fundamental limitations
of documentation-based government identity approaches.
First, they cannot achieve security or equality to any degree
without privacy-invasive tests and comparisons with documented records,
and hence necessarily fail to be privacy-preserving
even if might consider them plausibly securable.
Second, because identifying documents in either paper or digital form
are fundamentally just \emph{identity proxies} separate from a person
that can be lost, stolen, destroyed in a natural disaster or war,
or misappropriated through coercion,
they also fundamentally cannot enforce security or equality
without violating our goal of inclusion.
The myriad forms of people today who are excluded in practice
due substantially to the lack of (the right) documentation --
including undocumented migrants or homeless, refugees from disasters or wars,
those rendered stateless from lack of any provable citizenship, etc. --
illustrate the innumerable ways in which privacy-invasive identity approaches
exclude millions of real people whose only crime may have been to be unlucky.

\subsection{Biometric identity}
\label{sec:alts:biometric}

Even though government identity approaches
usually include biometrics as elements to varying degrees --
a person's photo on a passport or ID card being the standard baseline --
approaches that rely on biometrics primarily or even exclusively
for identification are worth examining in their own right.
The quintessential example of this approach
is India's Aadhaar program,
which has biometrically registered over a billion people
using iris and fingerprints~\cite{chaudhuri17aadhaar,abraham18state}.

The key attraction of this approach is that
a person's biometrics are essential characteristics of human physiology
that nearly everyone has and are quite unique to each individual.
Biometrics thus cannot readily be lost
like identity documents or mobile devices,
and cannot be forgotten
like passwords or answers to personal history questions.
Biometric technologies can also demonstrably be made
quite usable, efficient, and scalable:
just stand there and look here, place your fingers here, etc.

\subsubsection{Broad issues with biometric identity}

Biometric identity approaches face numerous technical, security,
and privacy challenges, however.
Even if people can't accidentally lose or forget biometrics,
they can be intentionally or unintentionally destroyed.
Fingerprints wear off from hard manual work.
People lose hands, arms, or eyes in accidents or violent conflicts.
Most biometrics also evolve gradually over time as a person ages.
Hackers have created wearable fake fingerprints,
contact lenses with iris patterns,
and even fake hands with embedded vein patterns,
regularly fooling even state-of-the-art biometric recognizers
with liveness detection.  \xxx{citations}
These factors and others likely contribute to
the increasing body of experiential evidence
that biometric identity systems are neither
as robust nor as inclusive
as they might at first seem~\cite{venkatanarayanan17enrolment,khera19aadhaar}.

Further, each electronic device used for biometric identity registration
or subsequent authentication --
and each official trusted with operating these devices
in registration and authentication processes --
represents a single point of failure or compromise.
These critical points
may be be exploitable either for identity theft,
improperly misappropriating the identities
of a legitimate victim~\cite{pritam18uttar},
or for Sybil attacks,
by synthesizing and registering multiple false identities
whose biometrics need not (and for the attacker preferably do not)
detectably match any existing user registered in the system
including their own real identity.
Large biometric identity databases have even been exploited
apparently for banal reasons
of unscrupulous business competition~\cite{venkatanarayanan17aadhaar}.

No matter how security-hardened these biometric devices might be,
there is unlikely to be a single trusted hardware technology
available today or in the forseeable future
secure enough to withstand a sustained attack
by a determined and resourceful adversary
focused on a particular device the adversary physically controls,
such as a biometric registration system that is stolen
or under the control of a compromised system administrator.
Thus, while biometric identity systems may well be secure enough
to detect or deter casual identity theft or fake-identity attacks,
their security against undetected attacks
by corrupt officials, resourceful criminal organizations,
or government spy agencies is far more doubtful.

\subsubsection{Biometric error rates and their implications}

Even when uncompromised, all biometric tests have nonzero
false-accept rates (FAR) and false-reject rates (FRR).
In state-of-the-art biometric technologies
of the type approved for use in Aadhaar for example,
these error rates tend to be in the 1-in-10,000 to 1-in-100,000 range,
which is usually more than adequate for biometric \emph{authentication} alone
but far more questionable as a basis for biometric \emph{identity}.
When a person unlocks their own mobile device
with a fingerprint or face recognition, for example,
this biometric authentication needs to compare the user present
only with one (or at most a few) templates stored on the device
representing the authorized user(s).

To implement biometric \emph{identity}, however --
including the \emph{deduplication} test needed
to detect and prevent Sybil attacks via duplicate registrations --
at registration time the user's biometric templates must be tested
for \emph{inequality} with all of the other (potentially billions of)
users already registered.
In this context,
even a state-of-the-art 1-in-100,000 false accept rate
for iris recognition
implies that a legitimate new registrant's iris pattern
may be expected to match falsely against 10,000 other irises
in Aadhaar's billion-user database.
Thus, a large-scale biometric identity scheme like Aadhaar
cannot rely on only one biometric but must rely on multiple biometrics --
like the two irises and ten fingerprints that Aadhaar uses --
and flag a potential duplicate only if
some threshold of templates in a new registration
match those in an existing record.

This potentially billion-fold increase in sensitivity to false positives
that biometric identity systems inevitably experience,
with respect to simple ``1-to-1'' biometric authentication,
correspondingly increases both the opportunities for fraudsters
and the exclusion threats to legitimate users.
Identity thieves may need to find a near-match
in only one biometric sample to find and exploit a false accept
against an existing real user's biometric identity.
Just as importantly for equality protection and Sybil resistance,
identity fraudsters might register false identities 
using plausible but randomly-synthesized biometric templates
of the kind regularly created for testing biometric technologies.
\xxx{citations}
Only one or two of the biometrics might match those of the real fraudster --
enough to pass subsequent simple authentication tests
of one particular fingerprint or iris, for example --
but few enough to remain under the duplicate alarm threshold
and to be plausibly deniable even even if duplication becomes suspected
for some other independent reason.
(``Of \emph{course} there are other identities
matching my left thumbprint in your billion-user database.
Your false-accept rate of 1-in-10,000 predicts
that there should be 100,000 such matches!'')

\subsubsection{Biometrics and privacy}

The dimension in which biometric identity
definitively fails our goals, of course, is privacy.
In contrast with 1-on-1 biometric authentication
against a template stored only within a mobile device to be unlocked,
for example,
the need for biometric identity registration to perform
an \emph{inequality} comparison between a new registrant's biometrics
and \emph{all} the previously-registered users
fundamentally requires that a centrally-queryable database
of all users' biometrics be built and maintained somehow, somewhere.
However it is created and managed,
this database becomes an extremely sensitive,
prime target for hacking and theft
by all manner of foreign governments and criminal organizations
wishing to track and identify people.

Furthermore, biometrics are, as they say,
passwords you can't change~\cite{schneier09tigers}.
Once a biometric database is compromised in any way for any reason,
everyone whose templates were stored in it
are permanently more vulnerable to surveillance or identity theft attacks
\emph{for life}.
Precisely because our they are the characteristics
most inextricably tied to our identities as physical beings,
biometrics are among the most sensitive and privacy-invasive
if overused or misused,
as amply illustrated in dystopian science fiction films
like \emph{Gattaca} or \emph{Minority Report}.
Despite the demonstrated appeal of biometrics
in terms of usability and scalability, therefore,
they definitely cannot meet our privacy goals for digital personhood.

\subsection{Self-sovereign identity}
\label{sec:alts:self-sov}

The key premise of self-sovereign identity~\cite{allen16path,muhle18survey}
is enabling users to collect government and third-party digital attestations 
of identity attributes such as name, age, citizenship, degrees, and so on,
and selectively reveal and prove those attributes to other parties on demand.
The ambition is to place people in charge of how their identities are used
and which aspects of their identity to reveal in a given interaction.

Self-sovereign identity
can certainly be useful for certain purposes that focus
on \emph{distinguishing between} people:
\eg, digitally verifying whether a job applicant
indeed has a claimed professional degree or certificate.
In some situations, self-sovereign identity may be privacy-preserving,
especially where the sole attribute to be revealed is a one-bit
yes/no or member/non-member test.
The quintessential example is proving one is old enough to drink legally
when entering a bar, without revealing anything else.

In scenarios calling for stronger verification
beyond boolean set-membership tests, however --
as generally required to prove an identity is ``official'' or ``unique'',
for example --
self-sovereign identity generally falls back on the traditional approaches
of requiring users to reveal the same kinds of rich, privacy-invasive attributes
that define government-issued and biometric identities.
If every business or organization that accepts a self-sovereign identity
for moderate- to high-trust purposes such as banking or online voting
must effectively demand a set of uniquely-identifying attributes
such as name, birthdate, birthplace, government-issued ID number, etc.,
then at least for these purposes self-sovereign identity is equivalent to,
and cannot fundamentally offer more privacy protection than,
conventional approaches to digital identity.

For digital democracy purposes,
the basic problem with self-sovereign identity
is that it still focuses on proving \emph{identity},
in terms of attributes that distinguish between and divide people,
rather than \emph{personhood},
in terms of empowering and protecting each real human being 
regardless of identity attributes.
Because digital democracy use-cases would require users to reveal
the same privacy-invasive, uniquely-identifying attributes
that government and biometric identities employ,
self-sovereign identity alone cannot offer the privacy protections we seek.
Since self-sovereign identity wallets on mobile devices may be lost or stolen,
users must be able to fall back on traditional processes and identity documents
to restore access,
leaving the fundamental exclusion challenges of identity approaches unsolved.
Finally, since any self-sovereign identity wallet and its attributes
may be used under coercion or bribery,
even with strong identity verification it is not ultimately suitable
for online voting or deliberation,
unless it is somehow enhanced with
coercion-resistance mechanisms of the kind
discussed above in \cref{sec:pop:coercion}.

\subsection{Proof of investment: work, stake, etc.}
\label{sec:alts:invest}

As with the identity approaches above,
for broad comparison purposes
it is worth examining Sybil-resistance mechanisms
used in popular permissionless cryptocurrencies,
even though these schemes generally make no attempt to achieve
the same goals as proof of personhood.
Nearly all of these schemes we can broadly classify
as \emph{proof of investment}:
anyone may participate,
but voting power and rewards are conferred
in proportion to each participant's amount of investment
in some activity or resource.

\subsubsection{Proof of work}
\label{sec:alts:pow}

Bitcoin~\cite{nakamoto08bitcoin}
was groundbreaking in that it created the first successful
\emph{permissionless} cryptocurrency,
allowing anyone in principle to join the network freely
and participate in consensus and community rewards
without prior identification or authorization.
The consensus algorithms driving Bitcoin
and most other deployed cryptocurrencies
are based on \emph{proof of work},
which had been previously proposed as a way to fight E-mail spam
and denial-of-service attacks~\cite{dwork92pricing,jakobsson99proofs}.

Proof of work is a cryptographic zero-knowledge proof technique
in which one party (the prover) convinces another party (the verifier)
that the prover expended a certain amount of computational effort
finding the proof, generally by solving cryptographic puzzles.
The verifier can check this proof quickly with minimal effort,
and in particular need not repeat all the prover's effort
finding the puzzle solution.
Bitcoin and many other permissionless cryptocurrencies 
use proof of work as a Sybil-resistance mechanism
by establishing a constant competition
between all first-class participants, or \emph{miners},
to solve proofs of work.
Each miner earns the right to participate in consensus, extend the blockchain,
and earn rewards,
in proportion to relative amount of work provably expended.

Proof of work's key strength and attraction as a Sybil-resistance mechanism
is privacy:
it does not require or demand any identity information
from either miners or end-users.
While Bitcoin's blockchain structure offers users only weak pseudonymity
because all transactions are publicly visible
on the blockchain~\cite{androulaki13evaluating,conti18survey},
subsequent permissionless cryptocurrencies offer
even stronger anonymity and transaction
privacy~\cite{sasson2014zerocash}.
Further, despite many subtle security issues being found
in Bitcoin and other cryptocurrencies
based on proof of work~\cite{eyal14majority,apostolaki16hijacking},
the overall security of permissionless consensus based on proof of work
still generally appears to have held up.

Given that Bitcoin's permissionless consensus was designed
specifically so that ``anyone'' could join and participate
at any time by mining,
we might expect permissionless cryptocurrencies to satisfy
our inclusion goal as well.
While it may still be true in a narrow sense
that anyone \emph{can} join and mine Bitcoin,
the development of specialized mining hardware
and the competitive landscape of mining economics
has led to the effective \emph{re-centralization} of most mining power
into the hands of a few entrenched specialists
with access to cheap power
and the latest mining hardware~\cite{vorick18state}.
For nearly anyone else,
first-class participation as a miner may still be possible
but is not \emph{economically} feasible:
one will pay far more in hardware and electricity
than one can hope to reap in rewards from participation.
Thus, proof of work's claim to inclusiveness has become questionable at best.

Finally, our goal
that proof of work fundamentally cannot (and does not attempt to) satisfy
is, of course, equality.
Because each proof-of-work miner
receives voting power and participation rewards
in proportion to invested computational effort,
and this computational effort costs real money,
proof of work clearly distributes influence and rewards to participants
with a rule much closer to a ``one dollar, one vote''
than a ``one person, one vote'' principle.

\subsubsection{Proof of stake}
\label{sec:alts:pos}

The high energy costs of the mining ``arms race''
that proof-of-work cryptocurrencies set up
has motivated intense interest in more energy-efficient alternatives,
one of the most popular being \emph{proof of stake}.
In this approach, participants must first obtain some existing cryptocurrency
in the proof-of-stake system --
either by being a founding member or buying some from an existing member --
and lock up or \emph{stake} these funds for some time period.
All stake-holders subsequently obtain voting power in permissionless consensus,
and participation rewards such as newly-minted cryptocurrency,
in proportion to their amount of stake.
Participants need not waste energy or any other physical resource,
but merely pay the opportunity cost
of not using their cryptocurrency for something else while it is staked.

Proof of stake protocols are certainly valuable alternatives to proof of work
for their energy savings alone.
They are also technically interesting and challenging to secure,
though these challenges generally appear
solvable~\cite{kiayias16ouroboros,gilad17algorand,badertscher19ouroboros}.
The main disadvantage from a perspective of our digital personhood goals
is again equality:
proof of stake is still a \emph{proof of investment} --
only a different form of investment than in proof of work --
and thus still operates in the ``one dollar, one vote'' paradigm.

Many other Sybil resistance schemes for permissionless cryptocurrencies
have been proposed,
such as proof of space~\cite{park18spacemint}
or even
proof of \emph{human} work~\cite{blocki16designing}.
These schemes generally retain the same fundamentally-unequal
\emph{proof of investment} character as proof of work or proof of space,
however,
giving proportionally more rewards and voting power
to anyone willing and able to invest more
in the appropriate resource or activity.

\subsection{Social trust networks}
\label{sec:alts:social}

We now explore a broad class of approaches to proof of personhood
that build on social networks and social trust principles.
We start with the PGP ``Web of Trust'' model
that first launched interest in this approach to digital identity,
then examine how direct social trust relationships
translate (or fail to translate) into plausible Sybil resistance properties.
Finally, explore the potentials and weaknesses
of Sybil resistance algorithms based on social graph analysis
and threshold identity verification tests.

\subsubsection{PGP's web-of-trust model}

Real human communities often rely on social trust in many ways:
\eg, social gossip as a source of information
and a means of judging its reliability,
\xxx{citations}
and word-of-mouth recommendations of (or warnings about)
people to hire for service tasks such as babysitting or repair.
Building on this basic aspect of human society,
PGP's \emph{web-of-trust} model~\cite{stallings95pgp}
first popularized the idea of building digital identity on social trust.

PGP's immediate goal was not to verify unique personhood
or resist Sybil attacks,
but instead merely to established social trust in mappings
between human-readable names and cryptographic public keys.
If Alice knows and already trusts Bob, for example,
and Bob introduces her online to someone named ``Charlie'',
Alice needs to know Charlie's correct public key
in order to authenticate and communicate with him securely.
Instead of just trusting any PGP public key labeled ``Charlie''
that she finds on the Internet --
which might well be an imposter trying to impersonate the real Charlie --
Alice can gain confidence 
that she has found the \emph{right} Charlie's public key
if Bob (or another trusted contact) has signed and attested to it.

Even when this ``web of trust'' is actually used --
which has been rare even among privacy activists --
this model establishes only
that Alice and Bob have a shared understanding
of what the correct public key is for their mutual contact ``Charlie.''
It does \emph{not} establish, for example,
that ``Charlie'' is that key-holder's real government-recognized name;
it might be merely a pseudonym that Charlie used
at the particular key-signing party at which Bob signed Charlie's key.
The web of trust similarly does not establish
that ``Charlie'' is the key-holder's \emph{only} pseudonym:
he or she might well hold many PGP keys under different pseudonyms
(``Dave,'' ``Eve,'' etc.).
He may even have obtained social trust signatures on them
at many PGP key-signing parties involving disjoint sets of participants.

This is merely the starting point for the issues we face
applying social trust to the goal of Sybil resistance for digital personhood:
social trust \emph{solves the wrong problem}.

\subsubsection{Social identity as a basis for Sybil resistance}

%\paragraph{Myth: users can build a reputation for not being a Sybil attacker.}

Many approaches to proof of personhood based on social trust
ask participants to verify their connections
and attest to their genuineness and uniqueness~\cite{shahaf20genuine}.
Some approaches ask users to vote on
whether they think an online identity is genuine,
as in HumanityDAO~\cite{rich19introducing}.
Some require users to \emph{stake} some form of currency
as a ``bet'' that those identities are not Sybils:
\eg, real money in Upala.\footnote{\url{https://upala-docs.readthedocs.io/}}
% XXX or identity ``health'' in Idena\footnote{\url{https://idena.io}}.
The expectation is generally that users should ``know'' their connections
well enough to be certain that they are not Sybil attackers.

When social trust works in practice,
it works by people acquiring a reputation
for reliably having some particular attribute or ability.
One acquires a social reputation as a dependable electrician,
sharp software developer, or talented musician
by doing those things, being observed doing them by friends or colleagues,
and being mentioned in others' social conversations as one who does them well.
One acquires a social reputation for kindness, or biting wit,
or an explosive temper,
by showing those personality traits,
and by those traits being discussed when one is absent.
Social trust works by propagating knowledge
of the \emph{presence} of certain abilities or character traits.

But the property of \emph{not} being a Sybil attacker --
\ie, of \emph{not} have any online personas
other than the ones a particular group of contacts knows about --
is an \emph{absence} rather than a \emph{presence}.
This simple fact places social trust schemes for Sybil resistance
in far more dubious territory.
How does one verifiably prove --
and earn a reputation among one's colleagues
or even one's closest friends --
of \emph{not} having any online alter egos unknown to them?
Almost everyone has alter egos:
different sides of their personalities or abilities
that they reveal only to certain (perhaps disjoint) subsets
of their friends and acquaintances.

Does the fact that none of your work colleagues
have witnessed you playing the piano
imply that they can, or should,
vouch that you can't play the piano,
\ie, that you have no alter ego as a pianist?
Obviously not: they may not have observed you playing the piano
simply because there is no piano at your workplace,
or you have no time to play it there.
Does the fact that none of your work colleagues
have observed you expressing interest in any sexual fetish
imply that you have no sexual fetish?
Obviously not: you're probably just (hopefully) well aware 
that your workplace is not the appropriate environment
in which to reveal or express that side of your personality.

It is hard to build and maintain a false social reputation
as a talented pianist
if no one in your social network has ever seen you play piano.
It is fundamentally much easier to lie to a group of friends
about \emph{not} being a talented pianist when you are one:
just don't play in their presence.
It is easy to prove convincingly
that you have an intimate relationship with someone:
just allow yourself to be seen kissing or holding hands with them.
It is fundamentally much harder to prove convincingly
that you \emph{don't} have an intimate relationship with anyone else:
just don't meet your secret lover in the presence
of your social friends or regular partner.
People manage this all the time.

In short,
social trust works to verify the \emph{presence} of personal attributes
because their presence is usually actually verifiable in some way.
Social trust does not work to verify the \emph{absence} of attributes,
including entire alter egos,
because they are easily and routinely hidden for many ordinary reasons.
The absence of attributes or alter egos is simply not socially verifiable,
other than by relying on a person's say-so and ``hoping'' they're not lying.

But this begs the question: should we even be \emph{asking} our friends
to attest or ``prove'' -- and perhaps lie about --
the absence of alternate identities unknown to us?

\subsubsection{Alter egos as a basic privacy right}

In practice,
a basic element of privacy
is the freedom to have alter egos:
the latitude to to express aspects of your interests, personality, or beliefs
in one social context that you're well aware may not be welcome
in another context.
People take on multiple personas and present different facets of themselves
in different contexts all the time:
\eg, at work with colleagues, versus at home with family,
versus with a group of friends sharing a particular common interest,
versus with a secret lover.
The fact that affairs or flings are so common --
which one might not disclose even to one's most trusted primary life partner
for a years, if ever --
makes it obvious how unrealistic and absurd the presumption is
that we can be certain about the \emph{absence} of another side
to a close friend or lover
based only on our \emph{absence of knowledge} of their having such a side.
And freedom-loving societies have come to recognize
that even if having an affair may break your fidelity vows,
that is none of the state's business --
and neither is it the business of
your work colleagues or nosy neighbors either,
except for those \emph{you} choose to confide in.

If having one or more secret alternate personas
represented by online identities
actually carried a strong negative social stigma,
then we might arguably hope that social attestation
might work to confirm the absence of Sybils for ``most'' people --
at least those with a strong moral compass
who are uncomfortable lying or just bad at it.
Even if having an online alter ego \emph{was} strongly stigmatised,
as is having a known predilection to go on a violent rampage,
social trust would still not detect everyone hiding that property --
as we can amply see in the regular news reports of mass murderers,
each of whose family and friends are ``shocked''
because the perpetrator always seemed like
such a nice, normal, upstanding person.
But in contrast with a strongly-stigmatized property like being a mass murderer,
having secret (online) personas in fact carries little to no social stigma,
because it is common and accepted for many ordinary reasons.

Who are people with multiple identities hurting, anyway?
The answer is no one, at least \emph{individually} --
only the social collective as a whole,
if each of those Sybil identities gets its own vote
and share in other benefits of society.
Sybil attack vulnerabilities create
a tragedy of the commons scenario that becomes readily apparent
only when attacks become severe enough
to undermine the equality, security, or legitimacy of a democracy
in obvious ways.

Alternate personas not only carry little negative social stigma,
but in some cases even a strong positive association.
The freedom to have multiple personas or alter egos is almost deified
in the cultural tradition of comic superheros,
nearly all of whom \emph{are} secret alter egos.
Must Clark Kent willfully lie to his friends and colleagues 
about having no other identities,
in order to serve society in his role as Superman?
It is almost taken for granted in democratic culture
that there are perfectly legitimate reasons
for one \emph{person} to maintain multiple \emph{identities} --
as long as the \emph{person}, whether Clark Kent or Superman,
casts only one vote.
Asking people to vouch that their friends have no online alter ego(s)
is privacy-invasive and disempowering in the basic presumption
that it is abnormal or unacceptable for a person
to have another identity representing an alter ego.
The very expectation that a person should have only one online \emph{identity},
in short, is actually a violation
of the freedoms we demand of digital \emph{personhood}.

It is therefore essentially immaterial
whether any of the proposed social trust schemes,
in which users are asked to verify
and vouch that their contacts are not Sybil identities,
could actually work securely.
Even if they did, they would fundamentally work against privacy,
effectively forbidding the normal human practice
of expressing multiple alter egos in different contexts in our lives,
some of which we may rightfully want or need to keep pseudonymous
and unlinked from others for legitimate privacy reasons.
Asking people to vouch or ``bet'' that their social contacts
have no other identities
effectively demands that friends, colleagues, and neighbors
to monitor each other constantly and snitch on them
at the slightest sign of having some previously-unknown personality facet,
just as in the worst historical surveillance states.

\com{
\subsubsection{Online verification ceremonies}
}

\subsubsection{Graph analysis for Sybil region detection}

Peer-to-peer networking research has produced a significant body of algorithms
that attempt to resist Sybil attacks
through structural analysis of a social graph.
SybilGuard~\cite{yu06sybilguard},
SybilLimit~\cite{yu08sybillimit},
SumUp~\cite{tran09sybil},
Wh\=anau~\cite{lesniewski-laas10whanau},
Gatekeeper~\cite{tran11optimal},
and
SybilRank~\cite{cao12aiding}
are just some examples.

Although these algorithms are technically interesting,
none of them satisfy our goals for digital personhood
for three fundamental reasons.
First, they are privacy-invasive by virtue of expecting everyone
to disclose all their social ties or trust relationships for analysis.
Second,
they do not detect Sybil \emph{identities} in general
but only Sybil \emph{regions} of a sufficient size --
only one narrow class of Sybil attack --
leaving other forms of attack wide-open to exploitation.
Third, social graph analysis cannot be effective
without also risking exclusion of legitimate users or groups
who might in fact be poorly-connected,
or for whatever reason
``look like'' part of a Sybil region to the algorithm.

Even if \emph{only} the structure of the social graph
is disclosed and available for analysis,
with no names or other identifying labels,
social graphs themselves tend to be unique enough
for effective re-identification using public
reference data~\cite{narayanan09de-anonymizing}.
We might hope to protect user privacy
by running the graph analysis algorithm under 
homomorphic encryption~\cite{gentry09fully}
or trusted hardware enclaves like Intel SGX.
But homomorphic encryption still incurs many orders of magnitude
higher computational costs than direct computation,
making its scalability and practicality for large social graphs doubtful,
and trusted hardware has regularly been found to have vulnerabilities
that allow supposedly-protected secrets
to be extracted~\cite{van2018foreshadow}.
Further, hiding the social graph would also make it inaccessible
for manual analysis to test, debug, or improve the analysis algorithms,
making it challenging for humans to investigate the algorithm's results
for false positives, false negatives, and newly-emerging
forms of attack for example.
The graph analysis algorithm would in effect become
another potentially-oppressive opaque governance-by-algorithm scheme,
offering little accountability or recourse
to users wrongly accused of being Sybils.

\paragraph{The Sybil-region movie plot:}

Just as a serious a limitation, however,
is that graph structure analysis algorithms
cannot actually detect individual Sybil \emph{identities}
but only, at best, Sybil \emph{regions} of sufficient size
that satisfy certain assumptions about the attacker's strategy.
These attack assumptions constitute
what Bruce Schneier might term a
\emph{movie-plot threat}~\cite{schneier05terrorists}:
one that ``captures the imagination'' --
in this case inspiring a whole sub-field of academic literature --
but which real-world attackers readily avoid
simply by adopting a different strategy.

\begin{figure}[t]
\includegraphics[width=0.90\columnwidth]{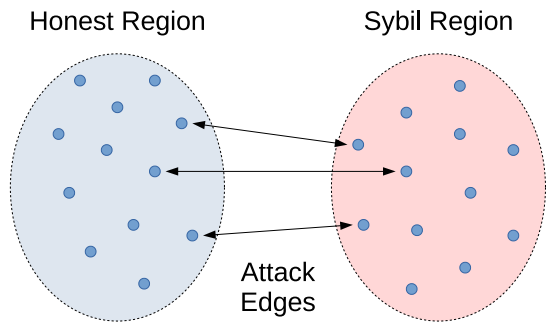}
\caption{The \emph{Sybil region} movie-plot threat
	assumed by graph-based Sybil resistance algorithms.}
\label{fig:sybil-region}
\end{figure}

In brief, graph-based Sybil-resistance algorithms
assume that Sybil attackers produce regions of the social graph
that look something like \cref{fig:sybil-region}.
A Sybil region consists of
a large number of Sybil identities that are densely connected internally,
but with a much smaller of \emph{attack edges},
or connections between the Sybil region
and ``honest'' identities of real users.
The intuition is that synthesizing any number of
internal nodes and edges within the Sybil region
is essentially ``free'' to the attacker,
while creating attack edges incurs some cost to the attacker.
For each attack edge,
the attacker must convince some real user
to accept a ``friend request'' to a fake Sybil identity.

The Sybil-region scenario, and graph-based defenses against it,
embody at least three dubious assumptions about the attacker's strategy:
(a) that the cost to the attacker
of creating attack edges is significant,
(b) that the attacker is unable or unwilling to pay this cost, and
(c) that the attacker wants to create one large Sybil region.
%as opposed to (for example)
%many attackers each wanting to create many small Sybil regions.
If any single one of these assumptions fails to hold in reality,
the protection these algorithms offer collapses completely.
In practice, \emph{all} of these assumptions probably fail to hold
or are easily circumvented by an attacker
who simply chooses not to follow the Sybil-region movie plot.

While attack edges probably do cost more to create
than purely-synthetic relationships among nodes within a Sybil region,
the presumption that this cost is \emph{significant} contradicts
practical experience with actual online social networks.
Real users and social bots alike
frequently ``follow'' or ``friend'' many other accounts indiscriminately --
and often automatically ``follow back'' any other account that follows them --
in order to build their follower counts
and ``influencer'' status~\cite{ferrara16rise}.
In effect,
the widespread use of friend or follower counts
as a reputation or influence metric
incentivizes behaviors that drive the effective cost of social connections --
and the real social trust they represent --
down towards zero.

Projects like Upala try to counter this social edge devaluation problem
by requiring identities to invest something of value in connections,
such as cryptocurrency costing real money.
Imposing such a financial barrier, however,
simply makes attack edges another form of ``proof of investment''
like proof of work or proof of stake,
as discussed above in \cref{sec:alts:invest}.
Any social connection cost high enough to deter even casual Sybil attackers
is exclusionary to people who can't afford the accepted price
of ``enough'' stake in even a few such connections.
And any social connection cost low enough for most people to afford
will be merely a modest ``cost of doing business''
for a wealthy Sybil attacker motivated to invest in many attack edges.
Thus, even if attack edges \emph{do} incur significant costs --
either in direct financial stake as in Upala,
or via indirect investments
such as creating sophisticated AI algorithms for social bot farming
or simply hiring real people to create plausible but fake online profiles --
these costs simply exclude genuine but financially-constrained users
while only modestly rate-limiting the capabilities of wealthy attackers.
The dominant paradigm remains ``one dollar, one vote''
rather than ``one person, one vote.''

The Sybil-region movie plot also implicitly presumes that
there is just one attacker,
whose goal is to create just one large Sybil region.
This is the online social network equivalent of a
James Bond villain,\footnote{\url{https://en.wikipedia.org/wiki/List_of_James_Bond_villains}}
the quintessential movie-plot threat.
In practice, however,
if each Sybil identity a person successfully obtains
confers proportionally more benefits,
such as votes in an election or universal basic income,
then \emph{all} participants have an incentive to obtain
as many Sybil identities as they can --
even if many poorer participants can ``afford'' only one or two Sybils.
Many participants might remain honest nevertheless purely for moral reasons,
but absent any significant barriers or deterrents,
many other participants may well succumb to the temptation
to cheat ``just a little''
while attempting to remain under the graph-analysis radar.

Further, if large, internally-dense
``Bond villain'' Sybil regions are readily detectable,
then smart attackers will instead simply create Sybils
that have few or no connections to each other,
mainly or exclusively relying on ``attack edge'' connections
to other real users.
An attacker adopting this strategy gives up the appealing prospect
of synthesizing a whole alternate universe of nearly-free
Sybil nodes and internal connections, of course,
and must now invest in a certain number of attack edges for each Sybil identity.
But this may again be simply a cost of doing business,
which rich attackers may be perfectly willing and able to pay,
to achieve non-financial objectives such as sowing misinformation
or ballot stuffing for example.

\begin{figure}[t]
\includegraphics[width=0.90\columnwidth]{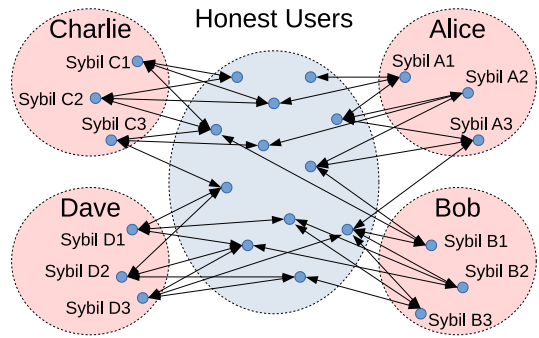}
\caption{Illustration of an alternative attack scenario
	that defeats graph-based Sybil resistance algorithms.
	Each of many attackers invests in a few Sybil identities,
	each of which relies mainly or exclusively on connections
	with other honest users instead of the attacker's other Sybils.}
\label{fig:sybil-forest}
\end{figure}

In summary, Figure~\ref{fig:sybil-forest} illustrates
an alternative attack scenario that may not capture the imagination
and inspire clever graph analysis algorithms
like the Sybil-region movie plot,
but is equally realistic in practice.
Instead of one Bond villain creating one large alternate Sybil universe,
many smaller attackers simply give in to their natural economic greed
by investing the effort and expense necessary
to create a few Sybil identities each,
connecting those identities mainly or exclusively
to other ``honest'' users rather than to their other Sybils.
Especially if these many small-scale Sybil attackers
take care to connect their Sybil identities
to \emph{disjoint} subsets of honest users,
they can also make it difficult for the detection of one Sybil identity
to lead to exposure of their other Sybil Sybil identities,
violating a key detectability assumption
in some Sybil-resistance schemes~\cite{shahaf20genuine}.

%com

\subsection{Threshold verification}
\label{sec:alts:verif}

Many proposed proof of personhood schemes
subject an online identity to some threshold test
of apparent ``genuineness''
in terms of representing a real human.
HumanityDAO~\cite{rich19introducing},
for example,
requires newly-proposed identities to receive
a sufficient threshold of ``yes'' votes
from existing users inspecting the identity.
BrightID asks users to build social connections
in regular online verification parties,
and achieve a threshold SybilRank score
to be verified~\cite{brightid20universal}.
Duniter requires users to have a threshold number
of social \emph{certifications}
and to be within a maximum social distance
from a distinguished \emph{referent member}~\cite{duniter18deep}.

\paragraph{Fakeability of profiles and verifications:}

Most of these schemes appear to have two significant weaknesses.
First, there is no obvious reason to believe
that automated virtual synthetic-identity attacks,
especially using deep fake techniques~\cite{chesney18deep},
cannot soon (or already) create convincing enough identities in bulk
to pass such ``genuineness'' threshold tests.
Automated techniques might soon be able to create fake profiles,
and even synthesized talking heads in online verifications,
that are just as convincing as real participants -- if not more so,
just as CAPTCHA-solving bots are already competitive with or surpassing
real humans' ability to solve CAPTCHAs~\cite{dzieza19why}.

In contrast,
our grounds for optimism that pseudonym parties
can remain secure against digital fakery at least for some time --
until convincing humanoid robots or biosynthetic clones
become readily available, for example --
is because pseudonym party transparency and security relies
not only on digital evidence
but also the direct observations or indirect attestations
of (ideally many) in-person eyewitnesses,
as discussed earlier in \cref{sec:pop}.

\paragraph{The cumulability of asynchronous verifications:}

The second key weakness is that
most threshold verification schemes for proof of personhood
require participants
to go through the verification process either once or periodically,
\emph{at a time of the participant's choosing}.
This property makes verifications for multiple Sybil identities
readily cumulable over time.
For example, a Sybil attacker might create one BrightID profile,
establish a threshold of social connections and get it verified
at one verification party;
then create another BrightID profile under a different pseudonym,
establish a new set of social connections for the new pseudonym
with a \emph{disjoint} set of other participants
at a \emph{different} verification party later,
and so on.
Even when verifications have an expiration, as in Duniter,
an attacker with some motivation
can still maintain many Sybil identities
while renewing each Sybil's certifications often enough
to to maintain its ``verified'' status.

There appears to be nothing these threshold verification protocols ask of users
that prevents one determined human
from completing exactly the same verification tasks
for two Sybil identities in succession,
if two different \emph{real} humans
could have performed the same verification tasks at the same respective times
on two otherwise-equivalent non-Sybil identities.
Without requiring some form of \emph{synchronized} task
that would require a Sybil attacker to be in two places at once,
as pseudonym parties rely on, 
it is not clear these threshold verification tests
have any way to distinguish between \emph{two humans} verifying real identities
and one \emph{time-shifting} human verifying Sybil identities.

\paragraph{Weaknesses to the elasticity of Sybil attackers:}

There are a few approaches to proof of personhood
that retain the idea
of assigning participants periodic mutual-verification tasks
at synchronized times,
and hence resist straightforward time-shifting attacks.
Encointer~\cite{brenzikofer19encointer}
retains even the \emph{in-person} element of pseudonym parties,
assigning small groups of participants to meet and verify each other
at randomly-assigned physical locations but at synchronized times,
so that a real person at one verification site cannot also be at another.
Idena moves these synchronized events online,
asking participants to challenge each other to
CAPTCHA-inspired ``FLIP tests''
to verify their humanness.\footnote{\url{https://idena.io}}
Pseudonym Pairs similarly assigns users to verify each other online
in pairs by interacting casually
in video chat sessions.\footnote{\url{https://panarchy.app/PseudonymPairs.pdf}}

Besides the risks of real-time digital fakery becoming sophisticated enough
to solve FLIP tests or otherwise trick real humans in online interactions,
as discussed above,
there is another significant potential weakness
that all of these approaches appear to have,
derived from their assignment of users to interact in small groups
(\eg, four per site in Encointer,
or pairs in Idena or Pseudonym Pairs).
These protocols appear to assume that a Sybil attacker is working alone
and thus has only one real human body to work with (their own).
This is not the case, unfortunately.
An attacker with motivation and funds might readily
use ``gig economy'' services
like Amazon Mechanical Turk\footnote{\url{https://www.mturk.com}}
to hire an elastic supply of real humans to perform tasks online --
such as participating in Idena or Pseudonym Pairs verifications --
under the attacker's central coordination.
Similarly, an attacker might use flexible ``in-person help'' services
like TaskRabbit\footnote{\url{https://www.taskrabbit.com}}
to obtain an elastic supply of participants
to attend Encointer meetings under the attacker's direction.
In either case,
we must keep in mind that these hired helpers
are not only elastic resources for the attacker
but also serve as replaceable parts, or \emph{minions}.
The attacker might send a different minion to represent, and be ``verified,''
in each successive event that one of his Sybil identities is asked to attend
in each cycle.

If such an attacker was always forced to hire as many minions in each cycle
as the number of Sybil identities the attacker wishes to maintain,
we would not consider this a successful Sybil attack:
the number of participating humans would be equal to the number of identities
in each cycle, independent of why each human participated.\footnote{
	This scenario would constitute
	a successful coercion attack, of course --
	relevant if the minions are hired to vote in support of the attacker,
	as discussed in \cref{sec:pop:coercion}.}
But unfortunately the attacker probably needs to hire
significantly fewer minions
than attacker-maintained Sybil identities in each cycle,
for at least two reasons.

First, these protocols cannot realistically expect genuine participants
to attend \emph{every} assigned meeting reliably,
whether it occurs in-person or online.
Instead, they can only reasonably expect an identity's human owner
to participate some threshold percentage of the time
in order to maintain ``verified'' status.
Further, these group verification protocols
only know about, and thus can only assign, \emph{identities} --
not the real humans purportedly behind them --
to verification groups,
without introducing privacy-invasive biometric tests and the like.
If a Sybil attacker knows that the system requires the holder of an identity
to show up to assigned meetings only 50\% of the time,
therefore,
the attacker need only hire one replaceable verification minion in each cycle
for every two Sybil identities the attacker wishes to maintain.
The attacker thus already has a $2\times$ Sybil advantage over honest users.

Pseudonym parties, in contrast,
do not even know about, let alone verify, \emph{identities}.
In particular, they never need to assign an identity to do anything.
Instead, \emph{real people} choose for themselves with their real bodies
which synchronized event to attend in each cycle, if any.
Attendees need not show up to any particular threshold number of cycles
in order to maintain ``verified status'':
there is no threshold verification.
The PoP tokens an attendee gets in each subsequent cycle
are completely independent and unlinkable.
Each \emph{person} gets a token in each cycle they attend,
and does not get a token in each cycle they miss.

The second key advantage a Sybil attacker can obtain,
in threshold protocols that rely on assigning identities to verification groups,
is more insidious because it may represent a small advantage initially
but allows the attacker to \emph{gain advantage} progressively over time
and eventually flood the system with Sybils.
Each time the protocol assigns identities to small verification groups,
there is some probability a given group will randomly contain
\emph{only} the attacker's Sybil identities.
Whenever the attacker ``gets lucky'' in such a meetup assignment,
the attacker need not hire or assign \emph{any} unique human minions
to that particular meetup time and place.
Since they are all virtual and attacker-controlled,
the two or a few Sybil identities can simply confirm that they all attended,
without actually doing anything.
If the protocol requires that the group record and publish evidence
(\eg, a video record)
that the meetup occurred,
then the attacker can either digitally forge that evidence at his leisure,
or employ only a few real minions to record many time-shifted ``meetups''
in succession.

Suppose the attacker initially invests in enough minions
to control 10\% of the total identities in the system, for example.
For each of the attacker's Sybil identities in each cycle,
the attacker experiences a roughly 10\% chance
of that Sybil ``getting lucky'' and being paired with another Sybil
in Idena or Pseudonym Pairs.\footnote{
	This probability is much lower with four-member Encointer groups
	(roughly $10\%^3 = .01\%$)
	but still may be non-negligible.}
If one of the participation benefits each Sybil identity receives
is a universal basic income in cryptocurrency~\cite{ford19money,zhang20popcoin},
for example,
and each identity needs to show up only 50\% of the time
to assigned meetups to remain verified as discussed above,
then the coordinated Sybil attacker gets both
a 50\% ``discount'' on the number of minions he must hire
due to the threshold requirement,
plus a further 10\% discount approximately from
attacker-dominated pairs.
The attacker can thus afford to pay each of his minions
slightly \emph{more}
than one identity's basic income is worth,
while still making over $2\times$ profit.

The attacker can now reinvest this profit
towards creating and maintaining more Sybil identities in subsequent cycles.
As the attacker's percentage of Sybil identities increases,
so does the percentage of assigned groups
that the attacker fully dominates,
and hence need not assign any minions to.
The attacker's advantage over honest users thus increases,
along with his effective hiring discount
since he needs to hire an even smaller number of minions each cycle,
leaving him with even more profit to invest in new Sybils,
and so on.
Once the attacker's ability to maintain Sybils grows
to control one-third of the total identities in the system in this scenario,
the attacker's minion hiring costs plateau at a constant.
He needs at most one real minion
to pair with every two honest identities in each cycle,
again because of the 50\% threshold requirement,
regardless of the number of Sybils the attacker creates.
The attacker now effectively controls the system completely,
and can claim any desired percentage of the system's benefits 
simply by creating more Sybils,
without further increasing the attacker's costs in real minions.

Increasing the size of the assigned groups
(\eg, from pairs to four members in Encointer)
exponentially decreases the attacker's \emph{initial} advantage
from completely-controlled groups for a given percentage of Sybil identities.
The use of larger groups does not affect the fact
that the attacker \emph{has} such a Sybil advantage, however,
which he can gradually increase over time
by reinvesting claimed benefits as described above.
Thus, unless the number of honest users constantly grows
faster than any attacker's Sybil identities,
it appears that any proof-of-personhood scheme of this form
that assigns identities to small groups for mutual verification
may eventually succumb to Sybil-attack takeovers, sooner or later.

\section{Conclusion}
\label{sec:concl}

Digital democracy cannot and will not exist securely
until it has a secure and usable
\emph{proof of personhood} foundation to build on.
This foundation must
robustly guarantee ``one person, one vote'' participation
while ensuring inclusion, equality, security, and privacy.
We have explored ways in which the previously-proposed idea
of pseudonym parties might be secured for small, medium, or large events,
and how they might be scaled geographically across many sites,
while ensuring that all organizing groups remain accountable to all others
through both digital evidence and direct cross-witnessing observations.
We have also explored some of the alternate approaches
commonly proposed as foundations for digital democracy,
and their weaknesses.
Government-issued, biometric, or self-sovereign identity approaches
can be Sybil-resilient only by being highly privacy-invasive.
Proof-of-investment methods, like proof of work and proof of stake,
offer privacy but not equality.
Proof of personhood approaches based on social networks,
or threshold verification mechanisms,
can potentially slow but cannot halt the creeping takeover of Sybil attackers.
Because the basic idea of proof-of-personhood
and all existing schemes are still new and immature, however,
much remains to be learned and new approaches no doubt await invention.

\subsection*{Acknowledgments}

The author wishes to thank
Jeff Allen,
Lucy Bernholz,
Linus Gasser,
Philipp Jovanovic,
H\'{e}l\`{e}ne Landemore,
Louis-Henri Merino,
Frederic Pont,
Rob Reich,
and
Haoqian Zhang
for numerous valuable discussions
contributing to the formation of ideas in this paper.

\com{
\section{XXX}

pseudonym~\cite{ford08nyms}

popcoin~\cite{borge17PoP}

chaum mix~\cite{chaum81untraceable} dining~\cite{chaum88dining}, blind~\cite{chaum82blind} cash~\cite{chaum88untraceable}

mixminioN~\cite{danezis03mixminion}
Tor~\cite{dingledine04tor}

blacklistable~\cite{tsang07blacklistable,tsang10blac,au12blacr}

anon cred~\cite{maheswaran13cryptobook,hohenberger14anonize,maheswaran16building}

accountable anonymity~\cite{diaz07accountable,corrigangibbs10dissent,feigenbaum11accountability,corrigangibbs13proactively,wolinsky12dissent,maheswaran13cryptobook,maheswaran16building,syta14security}

cosi~\cite{syta15decentralizing}

blockchain~\cite{kokoris16enhancing,nikitin17chainiac,kokoris17omniledger,syta16scalable}

sybil~\cite{yu08sybillimit,tran09sybil}

anonrep~\cite{zhai16anonrep}
}%com

\bibliographystyle{apalike}
\bibliography{bib/sec,bib/soc,bib/net,bib/os}

\end{document}